\documentstyle[twocolumn,prc,aps]{revtex}
\newcommand{\be}{\begin{eqnarray}}
\newcommand{\ee}{\end{eqnarray}}
\begin{document}

\twocolumn[\hsize\textwidth\columnwidth\hsize\csname @twocolumnfalse\endcsname
\title{New Excitation Mode of the Nucleon?}
\author{A.P.~Kobushkin}
\address{\it Laboratoire National Saturne, CEA/DSM CNRS/IN2P3,
F-91191 Gif-sur-Yvette Cedex, France\\
and\\
\it Bogolyubov Institute for Theoretical Physics,
National Academy of Sciences of Ukraine\\
252143, Kiev, Ukraine
}
\date{\today}
\maketitle

\begin{abstract}
We consider properties of narrow baryon states observed recently in
reaction $pp\to p\pi^+X$. Two lowest of them (with mass 1004 and 1044 MeV,
respectively) are stable against strong decay. Moreover we conclude that
they cannot decay to $\gamma N$ and thus this states are a kind of
metastable levels in quark system. The simplest decay channel is
assumed to be $2\gamma N$ and possible quark configuration with such
properties is discussed. New experiment for verification of status of
this states in $\gamma N$ collision is proposed.

\mbox{}
\end{abstract}

\vspace{0.1in}
]
\begin{narrowtext}

Recently three narrow bumps in missing mass spectra of the reaction
$pp\to p\pi +X$ with $m_X=$1004, 1044 and 1094 MeV have been observed
with good statistics by Tatischeff, Yonnet {\it et al.}
\cite{Tatischeff,TatiYonnet}.
The bump widths are probably in the range 4--15 MeV and are determined
by experimental resolution. The masses $m_X=$1004 and 1044 MeV are
below $m_{N}+m_{\pi}$ and thus this bumps can be interpreted as new
narrow baryon states stable against strong decay. A third bump is 19 MeV
above the $m_{N}+m_{\pi}$ threshold and allows an interpretation as a
possible cusp effect. In this paper we will denote
the possible new states by $N^{\prime}(1004),\ N^{\prime}(1044)$
and $N^{\prime}(1094)$, respectively. Of course physical widths
of the $N^{\prime}$ must be less than 4--15 MeV and cannot be
estimated from the data \cite{Tatischeff,TatiYonnet}.

The simplest decay channel of these states was naively expected to be
$\gamma N$. In this case the $N^{\prime}$ must contribute to Compton
scattering on the nucleon near $E_{\gamma}=$ 68, 112 and 169 MeV.
Nevertheless data analysis done by L'vov and Workman \cite{L'vov}
completely excludes $N^{\prime}$ excitations as intermediate states in
Compton scattering on the nucleon. The main aim of this Letter is to
show that the two lowest baryon states $N^{\prime}(1004)$ and
$N^{\prime}(1044)$ observed in the $pp\to p\pi^+X$ reaction
\cite{Tatischeff,TatiYonnet} can be considered as metastable levels
in nonstrange three quark system which cannot decay into one photon
channel. We propose a possible interpretation of this states as a member of
totally antisymmetric representation of the spin-flavor $SU(6)_{SF}$
group. Due to symmetry of spin-flavor wave function they cannot be
exited by one photon $\gamma N\to N^{\prime}$ or decay to $\gamma N$.
Actually the analysis \cite{L'vov} does not exclude {\it existence}
of the $N^{\prime}$, but only says that the transition
$\gamma N \leftrightarrow N^{\prime}$ is {\it forbidden} (at least for
the two lowest of them, $N^{\prime}(1004),\ N^{\prime}(1044)$).
Later on we will assume that the simplest decay channel of the
$N^{\prime}(1004),\ N^{\prime}(1044)$) is the two photon decay, $2\gamma N$.

From the poin of view of the general principles of the constituent quark
model there is a room for 3-quark states with the property of the
$N^{\prime}(1004),\ N^{\prime}(1044)$ if flavor-spin of their
wave function corresponds to \underline{20}-plet of the
$SU(6)_{FS}$. Feynman \cite{Feynman} mentioned that such states
cannot be transformed into a nucleon by operator acting on one
quark ({\it e.g.} electromagnetic current). Indeed the
\underline{20}-plet is totally antisymmetric representation of the
$SU(6)_{FS} $, [1$^3$]$_{FS}$, and one has to act on two different
quarks of it to create totally symmetrical representation [3]$_{FS}$
which includes the nucleon (Figure~1).

\includegraphics{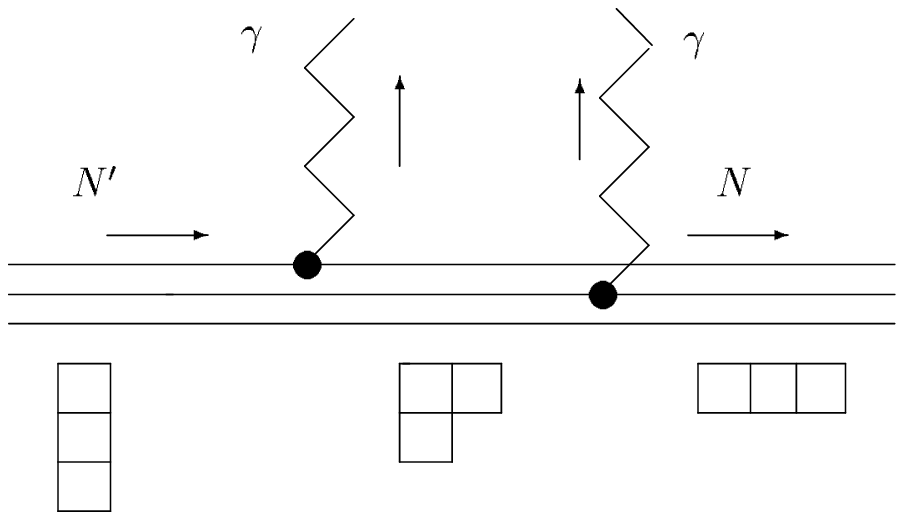}
\mbox{}

\vspace{4cm}
\mbox{}\\
FIG. 1. The quark diagram for
$N^{\prime}\to 2\gamma N$
decay. The first photon changes the spin-flavor symmetry from
[1$^3$]$_{SF}$ to [21]$_{SF}$; the second photon changes it as
[21]$_{SF}\to$[3]$_{SF}$.
\\

To excite such objects from the nucleon one needs (at least) two
steps for changing the spin-flavor symmetry. For example, in the
experiment \cite{Tatischeff,TatiYonnet} it was made as follows: at the
first step a resonance with mixed spin-flavor symmetry, [21]$_{SF}$,
is produced in $pp\to pN^{\star}$ collision. At the second step the
$N^{\star}$ decays on the narrow state and the pion,
$N^{\star}\to N^{\prime}\pi$ (Figure~2~a). This model predicts that
production cross section for the $N^{\prime}(1004),\ N^{\prime}(1044)$
must increase when the 4-momenta $p_1$, $p_2$ and $p_3$ of the
target, beam and final protons, respectively, are constrained by
\begin{equation}
\sqrt{(p_1+p_2-p_3)^2}\approx M_R
\label{-1}
\end{equation}
where $M_R$ is energy of one of resonances with [21]$_{SF}$ spin-flavor
wave function. The lowest of them are the negative parity resonances
$D_{13}$ and $S_{11}$ with average energy near 1520 MeV and a number
of resonances between 1600 and 1700 MeV.

The two step mechanism of the $N^{\prime}$ excitation is very similar
to laser pumping: first a level with energy higher that energy of
appropriate metastable state is excited and then quantum system transits
to the metastable state.

From their symmetry properties the $N^{\prime}$ states cannot be
excited at intermediate state of the Compton scattering
$\gamma N\to \gamma N$ at $E_{\gamma}=$ 68, 112 MeV. In turn, the two
steps production mechanism takes place in an ``inelasic Compton'' scattering
$\gamma N\to \gamma N^{\prime}$ which is described by the
same quark diagram as the  $N^{\prime}\to 2\gamma N$ decay, but with
inverse direction of quark lines and a line of one of the photons
(Figure~2~b). The appropriate photon energy must be
$E_{\gamma}=774$~MeV (which corresponds to $\sqrt{s}=1520$) or higher.

The \underline{20}-plet  consists of flavor octet and singlet with
$\frac{1}{2}$- and $\frac{3}{2}$-spin, respectively. The singlet has
nonzero strangeness and thus is excluded from our consideration.
To characterize an orbital wave function we will use the notations of
the translationally invariant shell model  \cite{Kurdyumov}
\begin{equation}
\Psi(\vec r_{12},\vec{\rho})=|N(\lambda,\mu)[f_X]LY_X>,
\label{0}
\end{equation}
where $N$ is the number of excitation quanta, $(\lambda,\mu)$
is the Elliot symbol determining the $SU(3)$ harmonic oscillator
multiplet, $[f_X]$ is the Young diagram for the spatial permutation
symmetry, $L$ is the total orbital momentum, $Y_X$ is the
Yamanuchi symbol specifying the basis vector of the representation
$[f_X]$ of the permutation group $S_3$, and $\vec r_{12}$
and $\vec{\rho}$ are Jacobi coordinates. For a totally
symmetrical spin-flavor state the Pauli principle requires wave
function  $|2(01)[1^3]1(123)>$, {\it i.e.} the
second excitation, $N=2$, with total orbital momentum $L=1$.
So one gets two desirable states with spin-parity-isospin
$J^{P}T=\frac{1}{2}^+\frac{1}{2},\ \frac{3}{2}^+\frac{1}{2}$,
respectively. In this case one of the bumps, say the 1094 MeV
bump (which is observed with less confidence), should have another
nature.

\includegraphics{fig2a.ps}

\includegraphics{fig2b.ps}

\vspace{7cm}
\mbox{}\\
FIG. 2. Examples of the $N^{\prime}$ excitation by the
two step mechanism: $pp\to p \pi+X$ (upper diagram) and inelasic
Compton scattering (lower diagram).\\

An important question of our model is how energy of 1600-1700 MeV
(typical mass of nucleon resonances from the second excitation) is
reduced to the energy of 1100 MeV and less. Of course it cannot be
related to a mechanism of a chiral constituent quark model by Glozman
and Riska \cite{Glozman} which explains low mass of the Roper resonance
and in any case it cannot be treated by perturbative calculation.
For example, naive application of the model \cite{Glozman} to our
object gives
\begin{equation}
m_{N^{\prime}}=3V_0 + 5\hbar \omega + 10 P_{11}.
\label{1}
\end{equation}
where $V_0=296.3$ MeV and $\hbar \omega=$157.4 MeV are parameters
of confining oscillator potential. The expectation value of pion
exchange potential $P_{11}$ was fitted to be positive, 45.2 MeV and
the masses of this states become very high. Nevertheless one can
assume existence of $LL$-coupling potential in three quark system
\begin{equation}
V_{LL}=\vec L^{(12)}\vec L^{(\rho)}v_{LL}(r_{12},\rho ),
\label{2}
\end{equation}
where $\vec L^{(12)}$ and $\vec L^{(\rho) }$ are angular momentum operators
corresponding to appropriate Jacobi coordinate. It does not contribute
to the energy of baryons
with $N=0$ and 1, as well as to the well established resonances with
$N=2$ and  the spatial permutation symmetry [3]$_X$ (for example to the
 Roper resonance). So the most important part of the baryon
spectroscopy \cite{Glozman,Glozman96} is not changed. In turn
for the $N^{\prime}$ states the angular momenta are $L=L^{(12)}=L^{(\rho)}
=1$; so
\begin{eqnarray}
&&<2(01)[1^3]1(123)|\vec L^{(12)}\vec L^{(\rho)}|2(01)[1^3]1(123)>=
\label{2.a}\\
&&
=\frac{1}{2}[L(L+1)-L^{(12)}(L^{(12)}+1)-L^{(\rho)}(L^{(\rho)}+1)]=-1
\nonumber
\end{eqnarray}
and the potential (\ref{2}) could strongly affect effective potential
for orbital motion of the quarks. At the moment one can say nothing
about the form of the potential $v_{LL}(r_{12},\rho )$.

The 40 MeV splitting between the two states may be explained by
spin-orbit interaction similar to that for usual baryons.

In conclusion,  we propose a model for narrow states observed
in \cite{Tatischeff,TatiYonnet}. According to their symmetry
properties they cannot be exited as intermediate state in Compton
scattering on the nucleon but can be produced in inelastic Compton
scattering $\gamma N\to \gamma N^{\prime}$.
Of course the $N^{\prime}$ states can be also exited in a reaction
where one of photons (or both photons) in the inelasic Compton scattering
is (are) replaced by a pion (pions)  at appropriate kinematic conditions.

Author acknowledges B.~Tatischeff and J.~Yonnet for important
information about their experiment and fruitful discussions, as well
as for O.P.~Pavlenko and E.A.~Strokovsky for reading manuscript.
He thanks authorities of Laboratoire National Saturne for warm
hospitality during his stay in Saclay.
The work  was supported in part by the Ukrainian State Foundation for
Funda\-mental Research $\cal N$o 2.5.1/41.

\end{narrowtext}

\end{document}